\documentclass[aps,epsf,preprint,showpacs]{revtex4}
\usepackage{latexsym,epsfig}

\begin{document}
\preprint{\vbox{\hbox{\tt gr-qc/0407106}}}
\title{Exactly Soluble Quantum Wormhole in Two Dimensions}
\author{Won Tae Kim}\email{wtkim@sogang.ac.kr}
\author{Edwin J. Son}\email{eddy@sogang.ac.kr}
\author{Myung Seok Yoon}\email{younms@sogang.ac.kr}
\affiliation{Department of Physics and Basic Science Research Institute,\\
 Sogang University, C.P.O. Box 1142, Seoul 100-611, Korea}

\date{\today}
\bigskip
\begin{abstract}
We are presenting a quantum traversable wormhole in an exactly soluble
two-dimensional model. This is different from previous works since
the exotic negative energy that supports the wormhole is generated from
the quantization of classical energy-momentum tensors. This
explicit illustration shows the quantum-mechanical energy can be
used as a candidate for the exotic source. As for the
traversability, after a particle travels through the wormhole,
the static initial wormhole geometry gets a back reaction which spoils
the wormhole structure.  However, it may still maintain the
initial structure along with the appropriate boundary condition.
\end{abstract}
\maketitle 

\section{Introduction}\label{sec:intro}
In a gravitational system, a black hole has the event horizon
and the curvature singularity, while 
a wormhole characterized by the throat is everywhere
regular. Interestingly, in the latter case, our universe 
can be connected to other universes in terms of its throat. 
However, the key ingredient in threading the two universes
is the violation of the energy theorem~\cite{mt,v,vkd}, which provides
a flaring-out condition near the throat.
It requires more or less an unusual source called
exotic matter. It is sometimes described by the negative energy
for simplicity. On the other hand, exactly soluble
classical wormhole models in two dimensions 
have been extensively studied by adding the negative energy source
in Refs.~\cite{swk,ks-D,hky}. However, the origin of the 
source is still unknown. Therefore, it will be interesting to study
some candidates of the exotic source.

We would like to present an exactly soluble 
traversable wormhole model without the classical exotic source.       
The exotic matter violating the energy theorem naturally 
arises from the quantization of real scalar fields so that the
quantum-mechanically induced energy may be a candidate~\cite{mty,nooo}. 
Motivated by these scenarios, we would like to explicitly show that
the necessary exotic source to support the wormhole
can be obtained from the quantum stress tensors.
The D-particle~\cite{pol-tasi} will be introduced as a test particle, 
whether it passes through the wormhole from our universe
to the other universe or not. The reason why we use the D-particle instead of
the usual particle is due to the exact solubility of our model.

In Sec.~\ref{sec:gen},  the geodesic of the D-particle and
its energy-momentum tensor are determined and the general solution of metric is
found without applying any boundary conditions. This solution
describes a wormhole or a black hole by the choice of the boundary condition.
We obtain the solution of the traversable wormhole by the proper
boundary condition in Sec.~\ref{sec:wh}. As a result, it will be given that
the formation of the wormhole is possible at the quantum
regime with the help of the quantum mechanically induced negative energy.
After the particle travels through the wormhole,
the static initial wormhole geometry gets a back reaction which spoils
the wormhole structure.  However, it is able to maintain the initial
wormhole structure along with the consistent vacuum state. 
In Sec.~\ref{sec:bh}, we choose another boundary condition to give
a black hole solution as a final state after the D-particle passes
through the wormhole. This is a different type of solution from the
conventional Russo-Susskind-Thorlacius(RST) model. Finally, in
Sec.~\ref{sec:diss}, discussions and summary are given.

\section{The RST Model Combined With A D-Particle}\label{sec:gen}
We now consider the Callan-Giddings-Harvey-Strominger(CGHS)
model~\cite{cghs} combined with the scalar fields and 
a D-particle, whose action is given by 
\begin{eqnarray}
  S_{\rm cl} &=& \frac{1}{2\pi} \int d^2x \sqrt{-g} e^{-2\phi} 
    \left[R + 4 (\nabla \phi)^2 + 4 \lambda^2 \right] + \frac{\epsilon}{2\pi}
    \int d^2 x \sqrt{-g}\sum_{i=1}^{N} \left[-\frac12(\nabla f_i)^2\right] \nonumber\\
    & & -  m\int d^2 x \int d\tau \delta^2
    (x-z(\tau)) e^{-\phi(x)} \sqrt{-g_{\mu\nu}(x)
    \frac{dz^\mu}{d\tau} \frac{dz^\nu}{d\tau}}, \label{act:cl}
\end{eqnarray}
where $g$, $\phi$, $\lambda^2$, and $m$ are a metric, a dilaton field,
a cosmological constant, and the mass of a D-particle,
respectively. The scalar fields $f_i$ are the real conformal fields
satisfying the energy condition for $\epsilon=1$. They are also the ghost fields
giving the negative energy density for $\epsilon=-1$. The D-particle
action instead of the conventional particle action  was introduced in
order to solve the model exactly without any approximations. For the
case of $\epsilon=0$ and $\epsilon=1$, the model describes a
collapsing D-black hole, which has been classically studied in
Ref.~\cite{ks-bh}. On the other hand, when $\epsilon=-1$, a
classical wormhole geometric structure appears due to the wrong sign
of the kinetic term in the action (\ref{act:cl}). This plays the role of the exotic
matter. A traversable wormhole is later obtained which the particle
can safely travel through. However, this may still open the 
problem concerning the origin of the classical negative energy density.

Returning back to our model, we now place the real scalar case of
$\epsilon=1$ in the action~(\ref{act:cl}). Of course, the wormhole solution in this
case does not exist because of the absence of the exotic source.
But, we semiclassically quantize the action by adding the one-loop
effective action of the real matter in the large N-limit as was done
in the RST model~\cite{rst},
\begin{eqnarray}
  S &=& \frac{1}{2\pi} \int d^2x \sqrt{-g} e^{-2\phi} 
    \left[R + 4 (\nabla \phi)^2 + 4 \lambda^2 \right] + \frac{1}{2\pi}
    \int d^2 x \sqrt{-g}\sum_{i=1}^{N} \left[ -\frac12(\nabla f_i)^2\right] \nonumber\\
    & &\! -  m\!\int\! d^2\! x\!\! \int\! d\tau \delta^2
    (x\!-\!z(\tau)) e^{\!-\phi(x)} \sqrt{-g_{\mu\nu}(x) 
    \frac{dz^\mu}{d\tau} \frac{dz^\nu}{d\tau}} 
    \!-\!\frac{\kappa}{2\pi}\! \int\!\! d^2\! x \sqrt{\!-\!g} 
    \left[ \frac14 R \frac{1}{\Box} R +\frac12 \phi R \right], \label{act:total} 
\end{eqnarray}
where $\kappa= (N-24)\hbar/12$. By introducing an auxiliary variable
$\eta(\tau)$, the Born-Infeld type action~\cite{pol-tasi,bi} for a
D-particle in the action (\ref{act:total}) can be rewritten as
\begin{equation}\label{act:D-e}
  S_{\rm D} = \frac12 \int d^2 x \int d\tau 
    \delta^2 (x-z(\tau)) \left[\eta^{-1}(\tau) g_{\mu\nu}(x) 
    \frac{dz^\mu}{d\tau} \frac{dz^\nu}{d\tau} - \eta(\tau) 
    m^2 e^{-2\phi(x)} \right],
\end{equation}
where the massless limit is well-defined.
In order to solve our model, we define new fields as~\cite{rst}
\begin{eqnarray}
  \chi &=& \sqrt{\kappa} \rho - \frac{\sqrt{\kappa}}{2} \phi +
    \frac{1}{\sqrt{\kappa}} e^{-2\phi}, \label{def:chi} \\
  \Omega &=& \frac{\sqrt{\kappa}}{2}\phi + \frac{1}{\sqrt{\kappa}}
    e^{-2\phi}, \label{def:Omega} \\
  \xi &=& m \int^\tau \eta(\tau) d\tau. \label{def:xi}
\end{eqnarray}
Subsequently, in the conformal gauge, $g_{+-} = - e^{2\rho}/2$, $g_{\pm\pm}=0$,
where $x^\pm = x^0 \pm x^1$, the action~(\ref{act:total}) takes
the form of
\begin{eqnarray}
  S &=& \frac{1}{\pi} \int d^2 x \left[ \partial_+ \Omega \partial_-
    \Omega -  \partial_+\chi \partial_- \chi + \lambda^2
    e^{\frac{2}{\sqrt{\kappa}} (\chi-\Omega)} + \frac12 \sum_{i=1}^{N} 
    \partial_+ f_i \partial_- f_i\right] \nonumber \\ 
  & & -m\int d^2 x \int d\xi \delta^2 (x-z(\xi))
    e^{-2\phi(x)}, \label{act:+-} 
\end{eqnarray}
with the constraints,
\begin{equation}\label{constr}
  \kappa t_\pm\! =\!  (\partial_\pm\Omega)^2 \!-\! (\partial_-\chi)^2 \!+\!
    \sqrt{\kappa}\partial_\pm^2 \chi\! + \frac12 \sum_{i=1}^{N}
    (\partial_\pm f_i)^2 \!+ T_{\pm\pm}^{\rm D},
\end{equation} 
where $t_\pm(x^\pm)$ reflects the nonlocality of the conformal anomaly
in the action (\ref{act:total}), which is fixed by some boundary
conditions. Then, the equations of motion may be found in
the action~(\ref{act:+-}) as 
\begin{eqnarray}
  \partial_+\partial_- \chi  + \frac{\lambda^2}{\sqrt{\kappa}}
    e^{\frac{2}{\sqrt{\kappa}} (\chi - \Omega)} &=& T_{+-}^{\rm D},
    \label{mot:chi} \\
  \partial_+\partial_- \Omega  + \frac{\lambda^2}{\sqrt{\kappa}}
    e^{\frac{2}{\sqrt{\kappa}} (\chi - \Omega)} &=& T_{+-}^{\rm D},
    \label{mot:Omega} \\
  \partial_{+} \partial_{-} f_i &=& 0, \label{mot:f} \\
  \frac{dz^+}{d\xi} \frac{dz^-}{d\xi} -
    e^{-2(\rho+\phi)} &=& 0, \label{mot:e+-} \\
  \frac{d^2 z^\pm}{d\xi^2} + 2\frac{\partial\rho}{\partial z^\pm} 
    \left( \frac{dz^\mp}{d\xi} \right)^2 &=& -2
    e^{-2(\rho+\phi)} \frac{\partial\phi}{\partial z^\mp}, \label{mot:z+-}
\end{eqnarray}
where the energy-momentum tensors for the D-particle are given by
\begin{eqnarray}
  T_{\pm\pm}^{\rm D} &=& \frac{\pi m}{2}  \int d\xi \delta^2(\!x-\!z)
    e^{2\rho} \left( \frac{dz^\mp}{d\xi}\right)^2, \label{T:D-same}
    \\
  T_{+-}^{\rm D} &=& \frac{\pi m}{2} \int d\xi \delta^2 (x-z)
    e^{-2\phi}. \label{T:D-diff}
\end{eqnarray}
From Eq.~(\ref{mot:f}), the solutions of the conformal matter
fields are simply $f^i = f_{+}^i(x^{+}) + f_{-}^i(x^{-})$. 
Combining Eqs.~(\ref{mot:chi}) and (\ref{mot:Omega}) yields
the reduced equation, $\partial_+\partial_- (\chi - \Omega) = 0$.
In the Kruskal gauge which fixes the residual spacetime symmetry,
a relation $\chi = \Omega$, {\it i.e.}, $\rho=\phi$, is obtained.
On the other hand, the einbein equation~(\ref{mot:e+-}) in the Kruskal
gauge is written as
\begin{equation}
  \label{mot:e-K}
  \frac{dz^+}{d\xi} \frac{dz^-}{d\xi} = e^{-4\rho(z)},
\end{equation}
and using Eq.~(\ref{mot:e-K}), the geodesic equation 
(\ref{mot:z+-}) becomes
\begin{equation}
  \label{mot:z+-1st}
  \frac{1}{A^\pm}\frac{dz^\pm}{d\xi} =  e^{-2\rho(z)}. 
\end{equation}
The particle geodesic is simply obtained as $z^+ = (A^+/A^-) (z^- + B)$, where
$A^\pm$ and $B$ are constants. Inserting Eq.~(\ref{mot:z+-1st}) into
Eq.~(\ref{mot:e-K}) yields the relation $A^+ A^- = 1$. 

For simplicity's sake, if we set $A =A^+$, the trajectory of the
D-particle is written as 
\begin{equation}
  \label{sol:D}
  z^+ = A^2 (z^- + B),
\end{equation}
describing the straight line in the Kruskal diagram. If the incident
D-particle starts from our universe, then it is effectively described
by the restriction, $A^2<1$. Note that the simple motion of
these particles is due to the exact solubility of our model. For
convenience, the energy-momentum tensors for the D-particle are
rewritten by substituting Eqs.~(\ref{mot:e-K})--(\ref{sol:D}) into
Eqs.~(\ref{T:D-same}) and  (\ref{T:D-diff}),
\begin{eqnarray}
  T_{++}^{\rm D} &=& \frac{\pi}{2} \frac{m}{A^3}
    \delta\left( \frac{x^+}{A^2} - x^- - B \right), \label{T:D++}  \\
  T_{+-}^{\rm D} &=& \frac{\pi}{2} \frac{m}{A}
    \delta\left( \frac{x^+}{A^2} - x^- - B \right), \label{T:D+-}  \\
  T_{--}^{\rm D} &=& \frac{\pi}{2} m A
    \delta\left( \frac{x^+}{A^2} - x^- - B \right), \label{T:D--}
\end{eqnarray}
where they are all singular along with the geodesic of the particle. 
By substituting Eq.~(\ref{T:D+-}) into Eq.~(\ref{mot:Omega}), we
get the geometric solution,
\begin{equation}
  \label{sol:Omega-a}
  \Omega = a_+ (x^+) \!+ a_- (x^-) - \lambda^2 x^+ x^- \! - \frac{\pi}{2}
     m A \left( \frac{x^+}{A^2} \!- x^- \!- B \right)
    \theta\left( \frac{x^+}{A^2} \!- x^- \!- B \right),
\end{equation}
where $\theta(x)=0$ for $x<0$ and $\theta(x)=1$ for $x>0$,
 and $a_\pm(x^\pm)$ should be determined 
by the constraints (\ref{constr}),
\begin{equation}
  \label{constr-a}
  \kappa t_\pm = \partial_\pm^2 a_\pm + \frac12 \sum_{i=1}^{N}
  (\partial_\pm f_\pm^i)^2.
\end{equation}
Integrating the constraints (\ref{constr-a}), we obtain the general
solution as
\begin{eqnarray}
  \Omega &=& - \frac{\lambda^2}{\sqrt{\kappa}} x^+ x^- + 
    \int^{x^+} \! dx^+ \int^{x^+}\! dx^+ \left[ \sqrt{\kappa}
    t_+ - \frac{1}{2\sqrt{\kappa}} \sum_{i=1}^{N} 
    (\partial_+ f_+^i)^2\right] \nonumber \\
  & & + \int^{x^-}\! dx^- \int^{x^-}\! dx^- \left[ \sqrt{\kappa} t_- 
    - \frac{1}{2\sqrt{\kappa}} \sum_{i=1}^{N} (\partial_- f_-^i)^2
    \right]   \nonumber \\
  & &- \frac{\pi}{2\sqrt{\kappa}} m A \left( \frac{x^+}{A^2} 
    - x^- - B \right) \theta\left( \frac{x^+}{A^2} - x^- - B \right) +
    C_+ x^+ + C_- x^- + D, \label{sol:Omega-gen}
\end{eqnarray}
where $C_\pm$ and $D$ are the constants of integration. Note that there are 
two large kinds of geometric solutions in our model. The first one is the
well-known RST black hole solution, of which asymptotic geometric structure is 
Minkowskian. We know it is given by the boundary condition of no incoming quantum
radiation. This incoming radiation is calculated by
\begin{equation}
  \label{rad}
  <T_{\pm\pm}^f> = \kappa [ \partial_\pm^2 \rho - (\partial_\pm
  \rho)^2 - t_\pm].
\end{equation}
The boundary conditions require $<T_{\pm\pm}^f> = 0 $ at $x^\mp
\rightarrow -\infty $ so that $t_\pm = 1/4(x^\pm)^2$~\cite{cghs}. 
The time-dependent solution may be found by patching the linear
dilaton vacuum, and the black hole across an infall-line~\cite{rst}:
\begin{eqnarray}
  \Omega &=& -\frac{\lambda^2}{\sqrt{\kappa}} x^+ x^- -
  \frac{\sqrt{\kappa}}{4} \ln (-\lambda^2 x^+ x^-) -
  \frac{M}{\lambda\sqrt{\kappa}x_0^2}(x^+ - x_0^+) \theta(x^+ - x_0^+)
  \nonumber \\
  & & -\frac{\pi mA}{2\sqrt{\kappa}} \left(\frac{x^+}{A^2}-x^- - B
  \right) \theta\left(\frac{x^+}{A^2}-x^- - B \right), \label{sol:bh}
\end{eqnarray}
where $\frac12 \sum_{i=1}^{N} \partial_+ f_i \partial_- f_i =
{M}/{(\lambda x_0^+)} \delta(x^+ - x_0^+)$ and $M >0$ is the energy
carried by the incoming shock wave.

\section{Traversable Wormhole From The Quantum Source}\label{sec:wh}
We would like to construct the wormhole solution
in the quantized theory by imposing a different boundary
condition from the previous black hole case. 
It means that in our soluble model, the past and the future horizon
curves are coincident with each other at the throat.  
In particular, the static wormhole appears at $x^+=x^-$.
The apparent horizon curves are also given by the definition,
\begin{eqnarray}
  0 &=& \partial_+ \Omega = - \frac{\lambda^2}{\sqrt{\kappa}} x^-
  + \int^{x^+} dx^+ \left[ \sqrt{\kappa} t_+ -
  \frac{1}{2\sqrt{\kappa}} \sum_{i=1}^{N} (\partial_+ f_+^i)^2
  \right] \nonumber\\
  & & \qquad \quad\ - \frac{\pi m/ A}{2\sqrt{\kappa}} \theta\left( \frac{x^+}{A^2} - x^-
  - B \right) + C_+, \label{hor:+} \\
  0 &=& \partial_- \Omega = - \frac{\lambda^2}{\sqrt{\kappa}} x^+
  + \int^{x^-} dx^- \left[ \sqrt{\kappa} t_- -
  \frac{1}{2\sqrt{\kappa}} \sum_{i=1}^{N} (\partial_- f_-^i)^2
  \right] \nonumber\\
  & & \qquad \quad\ + \frac{\pi m A}{2\sqrt{\kappa}} \theta\left( \frac{x^+}{A^2} - x^-
  - B \right) + C_-. \label{hor:-}
\end{eqnarray}
From the boundary condition of the static wormhole
geometry at the asymptotic past time, the unknowns $C_\pm$ and
$t_\pm$ are completely fixed as 
\begin{equation}
  \label{bc}
  C_\pm = \lambda^2 x_1, \qquad t_\pm =  \frac{\lambda^2}{\kappa},  
\end{equation}
where $x_1$ is the coordinate just before the infalling particle appears.
Note that we redefined the constant $D$ as
\begin{equation}
  \label{const:D}
  D = \frac{M}{\lambda} + \frac{\sqrt{\kappa}}{4}\left(1 - \ln
     \frac{\kappa}{4} \right) - \frac{\lambda^2}{\sqrt{\kappa}} x_1^2,
\end{equation}
for convenience. The constant $M>0$ was chosen 
from singularity-free condition of the curvature, which
will be discussed later.

\begin{figure}[tbp]
   \begin{center}
   \leavevmode
   \epsfxsize=0.5\textwidth
   \epsfbox{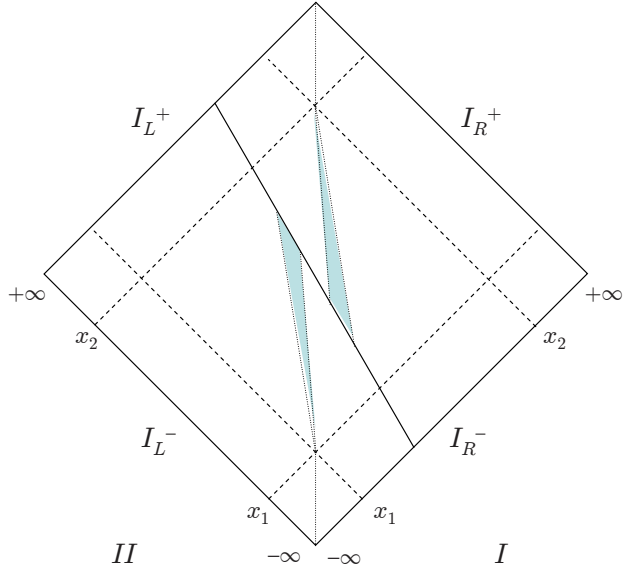}
   \end{center}
   \caption{A test D-particle passes through the
   wormhole. The geometry of the initial wormhole is perturbed by the
   infalling D-particle. However, it is recovered by the selection of
     the new vacuum state described by $t_\pm$.}
   \label{fig:wh}
\end{figure}

If the D-particle travels through the static wormhole, then the
spacetime is perturbed by the backreaction of the geometry. 
Now, we require that the initial wormhole structure be 
recovered after travelling at the later time, specifically, $x^\pm =
x_1$. This is easily realized by modifying the function $t_\pm$ in
Eq.~(\ref{bc}) as 
\begin{equation}
  \label{t+-}
  t_\pm = \frac{\lambda^2}{\kappa}[ 1 + 
    \beta_\pm (\theta(x^\pm - x_1) - \theta(x^\pm - x_2))],
\end{equation}
where the constants $\beta_{\pm}$ are chosen as,
\begin{equation}
  \label{beta}
  \beta_+ = \frac{\pi m/A}{2\lambda^2(x_2 - x_1)},\qquad 
  \beta_- = - \frac{\pi mA}{2\lambda^2(x_2 - x_1)}.
\end{equation}
They come from the static wormhole boundary condition of
the coincidence of the past and future horizons, 
\begin{eqnarray}
  0 &=& \partial_+ \Omega = \lambda^2 (x^+ - x^-) - \lambda^2 \beta_+ 
  (x_1 - x_2) - \frac{\pi m /A}{2}, \label{hor:beta+} \\
  0 &=& \partial_- \Omega = -\lambda^2 (x^+ - x^-) - \lambda^2 \beta_-
  (x_1 - x_2) + \frac{\pi m A}{2}, \label{hor:beta-}
\end{eqnarray}
at $x^\pm >  x_2$. 
Note that $x^\pm=x_1$ is the splitting point of the two horizons
caused by the incident particle and $x^\pm=x_2$ is the point where the
split horizons rejoin after the incident particle(Fig.~\ref{fig:wh}).
The incident travelling particle is defined between
$x_1$ and $x_2$, and it perturbes the wormhole geometry in this region.
The disturbed geometry is eventually stabilized through
the appropriately chosen $t_\pm$ in Eq.~(\ref{t+-}).
We are now able to figure out the exact time-dependent wormhole solution,  
\begin{eqnarray}
  \Omega &=& \frac{M}{\lambda} + \frac{\sqrt{\kappa}}{4} \left( 1 -
    \ln \frac{\kappa}{4} \right) + \frac{\lambda^2}{2\sqrt{\kappa}} 
    (x^+ - x^- )^2 + \frac{\lambda^2}{2\sqrt{\kappa}} \left[
    \beta_+ (x^+ - x_1)^2 \theta(x^+ - x_1)\right. \nonumber \\ 
  & & + \beta_- (x^- - x_1)^2 \theta(x^- - x_1) - \beta_+ (x^+ -
    x_2)^2 \theta(x^+ - x_2) \nonumber \\
  & & \left. - \beta_- (x^- - x_2)^2 \theta(x^- - x_2) \right]
   - \frac{\pi mA}{2\sqrt{\kappa}} \left( \frac{x^+}{A^2} 
    - x^- - B \right) \theta\left( \frac{x^+}{A^2} - x^- - B \right), \label{sol:metric}
\end{eqnarray}
which naturally yields the static wormhole, 
\begin{equation}
  \Omega = \frac{M'}{\lambda} + \frac{\sqrt{\kappa}}{4} \left( 1- \ln
  \frac{\kappa}{4} \right) + \frac{\lambda^2}{2\sqrt{\kappa}} (x^+ -
  x^-)^2  \label{sol:static}  
\end{equation}
satisfying the boundary conditions at $x^\pm < x_1$ and $x^\pm > x_2$. 
The new constant $M'$ is defined as
\begin{equation}
  \label{sol:M}
  M' = M +  \frac{\pi \lambda mA}{2\sqrt{\kappa}} \left[ B
    - \frac{(1-A^2)(x_2 + x_1)}{2A^2} \right]\delta,
\end{equation}
where $\delta$ is $0$ for $x^\pm < x_1$ and $1$ for $x^\pm > x_2$.
To make $M'$ positive definite, it should be 
\begin{equation}
  \label{avg}
  \frac{x_1 + x_2}{2} < \frac{BA^2}{1-A^2}.
\end{equation}

The constants $M$ and $M'$ characterize the sizes of the initial and
final wormhole throat, respectively.
In connection with the wormhole throat, $e^{-2\phi}$ 
is assumed to be analogously related to the higher-dimensional radial
coordinate~\cite{strominger}. It can be used to check whether the
wormhole is closed or not. Its radial size is defined by
\begin{equation}
  \label{throat}
  r^2 = \frac{e^{-2\phi_r}}{\lambda^2} > \frac{\kappa}{4\lambda^2},
\end{equation}
where $\phi_r$ is found in
\begin{equation}
 \label{throat:phi}
  \frac{\sqrt{\kappa}}{2} \phi_r + \frac{1}{\sqrt{\kappa}}
    e^{-2\phi_r} 
   = \frac{M'}{\lambda} + \frac{\sqrt{\kappa}}{4} \left(1-\ln
    \frac{\kappa}{4} \right),
\end{equation}
for the two regions of static wormholes, $x^\pm < x_1$ and $x^\pm
>x_2$. The size of the throat is larger than $\kappa/(4\lambda^2)$
from Eq.~(\ref{throat:phi}). This means the minimal size exists
even for $M=0$ and $M'=0$. In the quantum mechanical sense, as seen from the Planck
constant in Eq.~(\ref{act:total}), the static wormhole is always open
due to the quantum correction.

Compared to the previous black hole case, there 
should be one more constraint which is nothing but the regularity
condition. In this model, a curvature singularity may appear
at $d\Omega / d\phi = 0$ since $R={8e^{-2\phi}}/{\Omega'}
[\partial_+\partial_-\Omega - (\Omega''/\Omega') \partial_+
\Omega \partial_- \Omega]$, where $'$ denotes a derivative with respect
to $\phi$, and the singularity curve is given by $\Omega(x_+, x_-) =
{\sqrt{\kappa}}/{4}(1-\ln \kappa/{4})$. For the first time, the
singularity curves at the regions of the static wormholes are given by 
\begin{eqnarray}
  & & (x^+ - x^-)^2 + \frac{2M\sqrt{\kappa}}{\lambda^3} = 0, 
  \qquad\qquad\qquad\qquad\qquad\qquad\qquad\qquad\ \ 
 \ x^\pm < x_1
  \label{sing:x_1} \\
  & & (x^+ - x^-)^2 + \frac{2M\sqrt{\kappa}}{\lambda^3}  
    + \frac{\pi m A}{\lambda^2}\left[B- \frac{(1-A^2)(x_2 +
    x_1)}{2A^2} \right]= 0, \qquad \ x^\pm > x_2 \label{sing:x_2}
\end{eqnarray}
where $M$ should be positive for having no singularity from
Eq.~(\ref{sing:x_1}) at $x^\pm < x_1$. 
The spacetime at $x^\pm > x_2$
is regular as far as Eq.~(\ref{avg}) is satisfied. Next, to examine the 
singularity at $x_1 < x^\pm < x_2$, we assume that the case of $x_2 - x_1
= \pi mA / [2\lambda^2(1-A^2)]$, which gives the singularity curve as 
\begin{equation}
  \label{sing:split}
  a\left(x^+ \!- \!\frac{x^-}{a} \!+\! \frac{b}{2a}\right)^2 \!+\!
    \frac{2M\sqrt{\kappa}}{\lambda^3} \!+\! \frac{\pi m
    A}{\lambda^2}\left[B- \frac{(1\!-\!A^2)(x_2 \!+\!  x_1)}{2A^2} \right]
    \theta\left(\frac{x^+}{A^2} \!- \!x^-\! -\! B\right) = 0,
\end{equation}
where $a=1+\beta_+ > 0$ and $b = - 2\beta_+ [x_1 + (x_2-x_1)
\theta(x^+/A^2 - x^- - B)]$. This equation of the singularity curve
has no roots as long as $M >0$ and $M'>0$. Therefore the intermediate
spacetime is also regular. Our calculation was based on the very
restricted case rather than on general grounds because we wanted
to show the possibility avoiding the curvature singularity.       

\section{Transition From A Wormhole To A Black Hole}\label{sec:bh}
We have shown that there are two kinds of solutions in the quantized
theory. The first one is the well-known RST black hole solution, and
the second is the present dynamical wormhole solution. In the latter
case, the final state is the same with the initial wormhole, whose
size is a little bit different from that of the initial one. If this
is the case, one might ask whether the end state of our
wormhole can be the black hole solution or not. To patch a black hole
solution at $x^\pm =x_3 > x_2$, we should consider the appropriate
boundary conditions for the black hole. Since there is no incoming
quantum radiation, $<T_{\pm\pm}^f>=0$ at $x^\pm \rightarrow -\infty$
for $x^\pm > x_3$, we set $t_\pm=0$. So, the consistent boundary
condition gives
\begin{equation}
  \label{bh:t+-}
    t_\pm = \frac{\lambda^2}{\kappa}[ 1 + 
    \beta_\pm (\theta(x^\pm - x_1^\pm) - \theta(x^\pm - x_2^\pm))] -
    \frac{\lambda^2}{\kappa} \theta(x^\pm - x_3),
\end{equation}
which yields the solution by using Eq.~(\ref{sol:Omega-gen}),
\begin{eqnarray}
  \Omega\! &=&\! \frac{M}{\lambda} + \frac{\sqrt{\kappa}}{4} \left( 1 -
    \ln \frac{\kappa}{4} \right) + \frac{\lambda^2}{2\sqrt{\kappa}} 
    (x^+ - x^- )^2 + \frac{\lambda^2}{2\sqrt{\kappa}} \left[
    \beta_+ (x^+ - x_1)^2 \theta(x^+ - x_1)\right. \nonumber \\ 
  & & \!+ \beta_- (x^- - x_1)^2 \theta(x^- - x_1) - \beta_+ (x^+ -
    x_2)^2 \theta(x^+ - x_2) - \beta_- (x^- - x_2)^2 \theta(x^- - x_2)
    \nonumber \\
  & &\left. \!\!- (x^+\!\!\! -\! x_3)^2 \theta(x^+\!\! \!-\! x_3)\!
    -\!  (x^-\!\!\! -\! x_3)^2 \theta(x^-\!\!\! -\! x_3)  \right]\!
   \!- \!\frac{\pi mA}{2\sqrt{\kappa}} \!\!\left( \frac{x^+\!}{A^2} 
    \!-\! x^- \!\!-\! B \right)\! \theta\!\left( \frac{x^+\!}{A^2}\! -\! x^-
    \!\!\!-\! B \right), \label{wh-bh:metric} 
\end{eqnarray}
where the apparent horizon is $\partial_\pm \Omega = \lambda^2(-x^\pm + x_3)
/ \sqrt{\kappa} =0$. The patched black hole is unfortunately different
from the previous RST one since the boundary condition $t_\pm$ is
different from that of the RST model.  Therefore, the new type of the black
hole can be the final state of our dynamical wormhole. In this case,
the wormhole is no more traversable due to the size of the throat that
is shrunk to zero. Simultaneously, the infalling particle meets the
curvature singularity. 

\section{Discussion}\label{sec:diss}
Now, let us discuss the quantum energy-momentum tensors, which are of
relevance to the formation of the wormhole geometry. Essentially, the
exotic source in contrast to the normal matter satisfying the energy
condition should exist in order to support the wormhole structure. In
our model, the negative energy source has been obtained by the
quantization of the conformal matter fields instead of introducing by
hand. To make it explicit, let us first consider the static wormhole
geometry before the infalling  D-particle, which is achieved by 
letting  $m \rightarrow 0$ in Eq.~(\ref{sol:metric}) as  
$\Omega = M/\lambda + \sqrt{\kappa}/4 (1- \ln (\kappa/4)) +
\lambda^2(x^+ - x^-)^2 / (2\sqrt{\kappa})$.
Especially for a weak coupling, $\Omega \approx
{e^{-2\phi}}/{\sqrt{\kappa}}$, we get the exotic source  
$<T_{\pm\pm}^f> \approx -\lambda^2$ at $x^\mp \rightarrow -
\infty$. It corresponds to the Casimir vacuum of a quantum state
violating the energy condition discussed in Ref.~\cite{mty}.
After the D-particle travels, the final geometric structure approaches
the locally static wormhole. In that case, the energy-momentum
tensors are similarly calculated as $<T_{\pm\pm}^f> \approx -\lambda^2
(1+\beta_\pm) $ at $x^\mp \rightarrow -\infty$.
Therefore, the quantum-mechanically induced energy is the exotic
source that supports the wormholes in our model. 
 
The final comment is in order. The exoticity for the wormhole
solution~(\ref{sol:static}) defined in Ref.~\cite{mt} can be easily
checked. For this purpose, the proper reference frame of a set of
observers who remain always at rest in the coordinate system is
introduced as ${\rm \bf e}_{\hat 0} = e^{-\rho} {\rm \bf e}_0$ and
${\rm \bf e}_{\hat 1} = e^{-\rho} {\rm \bf e}_1$. In this basis, the
metric locally looks like an Minkowskian, $ds^2 = - (d\hat x^0)^2 +
(d\hat x^1)^2$. The energy momentum tensors, $<T_{\mu\nu}^f>$ written
by ($\hat 0$, $\hat 0$)- and  ($\hat 1$, $\hat 1$)-components in this
frame are $ < T_{\hat 0\hat 0}^f> \approx 2\lambda^2
e^{-2\phi_r} (3+4 e^{-2\phi_r}/\kappa)/(1-4 e^{-2\phi_r} /\kappa)$ and
$<T_{\hat 1\hat 1}^f> = - 2\lambda^2 e^{-2\phi_r}$ near the throat,
$|x^+ - x^- | \rightarrow 0$, where $\phi_r$ satisfies
Eq.~(\ref{throat:phi}). The specific dimensionless function $\zeta$
defined as $\zeta = (-T_{\hat 0\hat 0} - T_{\hat 1\hat 1})/|T_{\hat
  0\hat 0}|>0$ characterizes the exoticity of matter~\cite{mt}. Now,
it reads as $\zeta \approx [4\lambda^2 e^{-2\phi_r} (e^{-2\phi_r} +
\kappa/4 )] / (e^{-2\phi_r} - \kappa/4) > 0 $ near the throat since
$e^{-2\phi_r}> \kappa/4$ in Eq.~(\ref{throat}).

In summary, we have studied how the D-particle can travel through the
wormhole in the two-dimensional dilaton gravity coupled to the
D-particle. The crucial key to the formation and 
maintenance of the wormhole is to set the appropriate vacuum in the
quantized theory, which corresponds to appropriate choice of $t_\pm$
in our model. As a result, we have shown that in a simplified model
calculation the quantum-mechanically induced
energy may be a candidate of the exotic source for the wormhole.

\begin{acknowledgments}
  This research was supported by the Sogang University Research Grants
  in 2004.
\end{acknowledgments}

\end{document}